\documentclass{svproc}
\usepackage{comment}
\usepackage{graphicx}%
\usepackage{multirow}%
\usepackage{amsmath,amssymb,amsfonts}%
\usepackage{mathrsfs}%
\usepackage[title]{appendix}%
\usepackage{xcolor}%
\usepackage{textcomp}%
\usepackage{manyfoot}%
\usepackage{booktabs}%
\usepackage{algorithm}%
\usepackage{algorithmicx}%
\usepackage{algpseudocode}%
\usepackage{listings}%
\usepackage{url}

\begin{document}
\mainmatter              % start of a contribution
\title{Out-of-Network Attention Dynamics on Bluesky}
\titlerunning{Out-of-Network Dynamics on Bluesky}  % abbreviated title (for running head)
\author{Andrea Failla\inst{1} \and Veronica Mesina\inst{1,2} \and
Giulio Rossetti \inst{1}}
\authorrunning{Andrea Failla et al.} % abbreviated author list (for running head)

\institute{Institute of Information Science and Technologies ``A. Faedo'' (ISTI),\\ National Research Council (CNR), Pisa, Italy\\
\email{\{name.surname\}@isti.cnr.it}
\and
Department of Computer Science, University of Pisa, Pisa, Italy}

\maketitle              % typeset the title of the contribution

\begin{abstract}
Personalized social media commonly relies on explicit follow graphs to shape what content users encounter; yet how much attention crosses ties they have not formed remains largely undocumented at scale. We study this question on Bluesky, a large decentralized microblogging platform whose default feed relies on a simple, reverse-chronological content recommender. We analyze $173$ million user--author interactions (likes, reposts, replies, and quotes) collected from a near-complete platform dump between February and September $2023$. We decompose each interaction by attention-path length (already followed, relayed by a followed account, reachable within two follow-hops, or beyond) and find that $74.5\%$ of interactions reach the user through an account they already follow. Measured by distance in the follow graph rather than by route, $80.6\%$ of interaction lands within two follow hops, far beyond the $22.4\%$ an expected-degree null predicts. We then characterize how exploration varies across users and over tenure. A broad-reaching minority generates three quarters of all exploratory activity, while aggregate declines in exploration with tenure mask three distinct individual trajectories. Finally, attention reaching beyond two hops converts into new follow ties at less than one third the rate of two-hop-local exploratory attention. Together, these results depict a platform where out-of-network exploration is substantial in volume but strongly constrained by network proximity and unlikely to translate into new social ties.

\end{abstract}
\section{Introduction}\label{sec1}

Most large social media platforms mediate content exposure through ranking algorithms whose operation is only partially observable from the outside.
This makes it difficult to disentangle two forces that can structure online attention: the social neighborhood users construct through explicit ties and the content that recommender systems choose to surface from within or beyond that neighborhood.
Bluesky offers an unusual setting in which to study this distinction.
Built on the open AT Protocol~\cite{kleppmann2024atproto}, it uses a default \emph{Following} feed that presents posts from followed accounts in reverse chronological order.
Other recommendation mechanisms exist, but they are user-selectable alternatives, and existing large-scale studies suggest that their uptake remained limited relative to the platform's user base~\cite{failla2024m,quellebovet2024topology}.
The Bluesky follow graph therefore provides an unusually meaningful structural baseline for asking how far users' attention extends beyond the social neighborhood they have explicitly constructed.
\\ \ \\
In this work, we use \emph{attention} to refer to the behavioral traces through which users engage with content.
Following the literature on collective attention, likes, reposts, replies, and quotes provide observable signals of where users direct their attention~\cite{weng2012memes,lorenzspreen2019collective}.
This allows us to ask a basic structural question: \textit{when users engage with content, how much of that attention remains close to the accounts they already follow, and how much reaches beyond their existing network?}
The question matters because social ties are themselves selective.
Homophily, i.e., the tendency of similar individuals to connect, implies that the follow graph already constitutes a non-random information neighborhood \cite{mcpherson2001homophily}, while algorithmic curation can further concentrate exposure around existing interests and relationships~\cite{pariser2011filterbubble,cinelli2021echo}.
Understanding how much of that attention the follow graph itself accounts for, before attributing any narrowing to recommendation, is therefore important for separating the role of network structure from that of content-ranking mechanisms.
At the same time, attention that moves beyond one's immediate neighborhood may be especially consequential. Classic work on weak ties argues that boundary-spanning connections provide access to novel information and social circles~\cite{granovetter1973weakties}.
Out-of-network attention may play an analogous role: it exposes users to accounts outside their established neighborhood and may, when followed by tie formation, redraw the boundary of that neighborhood itself.
Despite this, out-of-network attention remains unexplored to date.
To address this gap, we organize the analysis around three research questions: \textbf{RQ1:} How much observable attention escapes the follow graph? \textbf{RQ2:} How is exploration distributed across users, and does it change as users spend more time on the platform? \textbf{RQ3:} Does attention that escapes the follow graph convert into new social ties?
\\ \ \\
Using a near-complete collection of 173M interactions by 1.1M users, we find that most of what users act on reaches them through accounts they already follow: 74.5\% of interactions arrive either from a followed author or by reshare from a followed account. 80.6\% of interaction lands within two follow hops, substantially above what produced by an expected-degree null model.
Exploration is nevertheless substantial in absolute volume and highly heterogeneous across users: a broad-reaching minority generates most exploratory activity, while aggregate declines with tenure conceal distinct upward, downward, and transient individual trajectories.
Finally, structural distance is associated with markedly lower tie formation: interactions beyond the two-hop neighborhood convert into follows at less than one third the rate of local exploratory interactions.\\
The rest of this work is organized as follows. Section~\ref{sec:relwork} reviews related work; Section~\ref{sec:methods} describes the data, operationalization, and analytical methods; Section~\ref{sec:results} presents the results; and Section~\ref{sec:discussion} discusses the findings, limitations, and directions for future work.

\section{Related Work}\label{sec:relwork}
\textbf{Network structure and attention.}
Online attention is shaped not only by platform design but also by users'
own social and informational preferences. Homophily leads individuals to form
ties with similar others~\cite{mcpherson2001homophily}, while selective
exposure and confirmation bias make people more likely to seek out and engage
with information consistent with their existing beliefs and interests
\cite{nickerson1998confirmation,garrett2009echo}. These mechanisms can produce
substantial self-selection into socially and informationally homogeneous
neighborhoods even before algorithmic ranking is considered. On social media,
such tendencies are reflected in the emergence of clustered and sometimes
ideologically segregated interaction patterns~\cite{cinelli2021echo}.
Most existing work, however, characterizes this concentration at the level of
communities, ideology, or information diffusion rather than measuring how far
individual acts of attention extend from a user's own follow neighborhood.
We take this structural distance itself as the object of study.
\\ \ \\
\textbf{Boundary-spanning interaction and tie formation.}
Attention outside an established social neighborhood is potentially important
because weak and bridging ties can connect otherwise distant parts of a network
and facilitate access to novel information~\cite{granovetter1973weakties,bakshy2012role,zhao2010weak}. Yet boundary-spanning contact does not
necessarily become a persistent structural relation. Work on evolving online
networks shows that tie formation is shaped by prior network position,
interaction, and mechanisms such as triadic closure
\cite{ameri2023strangers}. We therefore distinguish between
\emph{exploration} --- interaction with accounts outside the immediate follow
neighborhood --- and its subsequent conversion into a new follow tie. This
connects the literature on information reach with that on network evolution.
\\ \ \\
\textbf{Bluesky.}
Existing work on Bluesky has primarily characterized the platform's growth,
network structure, and forms of participation. Failla and Rossetti
\cite{failla2024m} provide a large-scale description of activity, users, and
the follower network; Quelle and Bovet~\cite{quellebovet2024topology} study
network topology, political structure, and the ecosystem of algorithmic feeds.
More recent work examines how Starter Packs affect onboarding and diffusion
\cite{failla2026starterpacks} and how participation differs in intensity and
behavioral style~\cite{mesina2026lowvocality}. These studies describe who uses
Bluesky, how they participate, and how platform mechanisms structure activity.
We instead focus on the spatial relation between interaction and the follow
graph.

\section{Methods}\label{sec:methods}

\noindent\textbf{Data Collection.} Data for our study are collected via the Bluesky API.
Bluesky stores user data in Personal Data Servers (PDSes), publicly accessible repositories containing the full history of a user's activity on the platform.
Starting from a sample of approximately $5$ million users~\cite{failla2024m}, we query their PDSes for all available activity records up to January $2025$.
The dataset includes multiple action types, among them follows, likes, posts, reposts, and blocks; for a complete description we refer the reader to the official documentation~\cite{atprotoDocs}, and to~\cite{kleppmann2024atproto} for an architectural overview of the protocol.
For our analysis to remain computationally tractable, we restrict it to the first 8 months of platform activity, from $2023$-$02$ to $2023$-$09$.
\\ \ \\
\noindent\textbf{Attention-Path Length.}
We operationalize attention as observed interaction.
The primary unit of analysis is a user--author interaction $d_{u,a}$: the interaction between user $u$ and a piece of content authored by $a$, over four types (likes, reposts, replies, quotes).
We argue that an interaction is sufficient evidence that $u$ attended to $a$'s content; on the other hand, it is not necessary evidence of attention's absence, since exposure a user does not act on (e.g., a scroll, a read) leaves no trace in this data.
Every quantity reported under this axis is therefore a lower bound on attention proper.
We measure that attention along the shortest observed path between a user and the content they act on.
Every interaction falls into exactly one of four tiers.
An interaction is \emph{in-network} if $u$ already follows $a$ at the time of interaction, or if $u=a$, and \emph{out-of-network} otherwise.
An out-of-network interaction is \emph{relay-mediated} if an account followed by $u$ reposted or quoted the target post $p$ before $d_{u,a}$, with both the follow and relay preceding the interaction.
This identifies a possible route to the post rather than its actual delivery mechanism; conversely, the absence of a relay does not rule out recommendation or search.
Out-of-network interactions without such a relay are \emph{exploratory}. We further distinguish exploratory interactions as \emph{two-hop local} when a directed follow path $u \rightarrow i \rightarrow a$ exists, and \emph{distant} otherwise.
\\ \ \\
\noindent\textbf{Clustering.} We characterize how exploratory behavior is distributed across users with two complementary clusterings, fit and validated by the procedure detailed in Appendix~\ref{app:clustering}. The first clusters users on the \emph{composition} of their exploration --- rate, breadth, and repeat contact --- to characterize what kind of explorer a user is. The second clusters each user's own eight-week exploration-rate \emph{trajectory}, $z$-scored to remove overall level, to characterize whether that behavior is rising, falling, or transient over their early tenure.
\\ \ \\
\noindent\textbf{Conversion Rate Analysis.} We treat a user--author pair's first follow after $t_0$, if any, as the outcome of interest and model it in two complementary ways.
Descriptively, we report the share of pairs converting by $7$, $15$, and $30$ days, overall and split by the scope of the pair's first contact (two-hop local vs.\ distant) and crossed with first-event type.
Adjusted, we fit pair-level logistic regressions of $30$-day conversion on first-event and user-covariate features (full specification in Appendix~\ref{app:model}); this isolates the association between scope and conversion net of the user characteristics we can measure, without claiming a causal effect of exploring further from one's network.

\section{Results}\label{sec:results}

\subsection{RQ1: How Much Attention Escapes the Follow Graph?}

Of $1.1$M users with at least one analyzed interaction, $575$K contribute an explored pair. {Fig.~\ref{fig:null} (top) decomposes the $173$M interactions by attention-path length: $62.3\%$ involve an already-followed author, $12.2\%$ are relay-mediated by a followed account, $18.3\%$ reach an author two follow-hops away, and $7.2\%$ extend beyond two hops.}
These tiers support two distinct summaries. First, $74.5\%$ of interactions are routed through an account the user already follows, either as author or relay. This concerns possible delivery routes, not graph distance: a relay does not imply a two-hop follow path to the original author.
{Second, considering follow-graph distance and excluding relays from the distance count, $80.6\%$ of interactions reach authors within two follow hops: $62.3\%$ at one hop and $18.3\%$ at two. Because two-hop reachability is evaluated only for non-relayed interactions, this is a lower bound: some relay-mediated interactions may also lie within two hops.
To assess whether this concentration exceeds what degree structure alone would imply, Fig.~\ref{fig:null} (bottom) compares the observed share with an expected-degree null preserving users' in- and out-degree in expectation. Only $22.4\%$ are within two hops under the null, a $58.4$-point gap that remains stable across time-separated cohorts (details and coverage in Appendix~\ref{app:null}).

\begin{figure}[!ht]
\centering
\includegraphics[width=\linewidth]{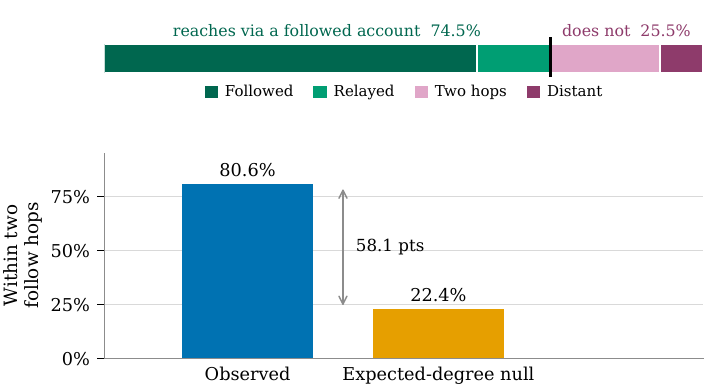}
\caption{{Top: the analyzed interactions decomposed by attention-path length ($107.8$M followed, $21.2$M relayed, $31.7$M two-hop local, $12.5$M distant) Bottom: the share of interaction landing within two follow hops  (i.e., the followed and two-hop-local tiers, relay excluded) against an expected-degree null.}}
\label{fig:null}
\end{figure}

\subsection{RQ2: How is Exploration Distributed Across Users, and Does it Change with Tenure?}\label{sec:tenure}

We first ask, at a coarse grain, whether the users who follow and interact most are also the ones whose exploratory attention stays closest to home. Ranking users into deciles by total interaction volume, and separately by out-degree (accounts followed), the first decile holds the tenth of users with the lowest value on that measure and the tenth decile the tenth with the highest.
Local share of exploratory interaction climbs from $20.1\%$ in the first activity decile to $74.3\%$ in the tenth, and from $3.3\%$ to $80.4\%$ across out-degree deciles: the users who interact and follow most are the ones whose exploration stays closest to home, a between-user gradient we return to with the clustering below.

We then move to two complementary clusterings of user behavior. \\
\noindent\textbf{Composition clustering.} We cluster users on what their exploration looks like -- exploration rate, breadth (distinct explored authors), and repeat contact per author, each standardized before clustering for the $246K$ users with at least ten explored events.
This yields $K=4$ clusters.
Each feature singles out a different cluster against a shared baseline.
Cluster 3, the $53.3\%$-of-users majority (``narrow, occasional, low-rate'' in Figure~\ref{fig:clusters1}), is low on all three features; each other cluster departs from that baseline on exactly one.
Breadth singles out the ``broad, occasional, low-rate'' cluster ($22.7\%$ of users, reaching $293.9$ authors on average, five to fourteen times as many as any other cluster); repeat contact singles out the ``narrow, repeated, low-rate'' cluster ($5.2\%$ of users, revisiting each explored author $5.2$ times on average against $1.4$--$1.8$ for the rest); exploration rate singles out the ``narrow, occasional, high-rate'' cluster ($18.8\%$ of users, $66.7\%$ of whose activity is exploratory) despite it reaching only $31.6$ authors on average, among the narrowest of the four.
We can see that breadth and exploration rate are not necessarily positively related: the highest-rate cluster is one of the narrowest-reaching, while the broadest-reaching cluster explores at a middling $29.7\%$ and seldom returns ($1.76$ events per author).
The broad, low-repeat cluster is $22.7\%$ of users but $74.4\%$ of all explored events, while the narrow, occasional majority contributes $9.1\%$: exploration volume is concentrated in users who reach widely and lightly, not in those who explore most intensively per interaction.

\begin{figure}[!ht]
\centering
\includegraphics[width=1.05\linewidth]{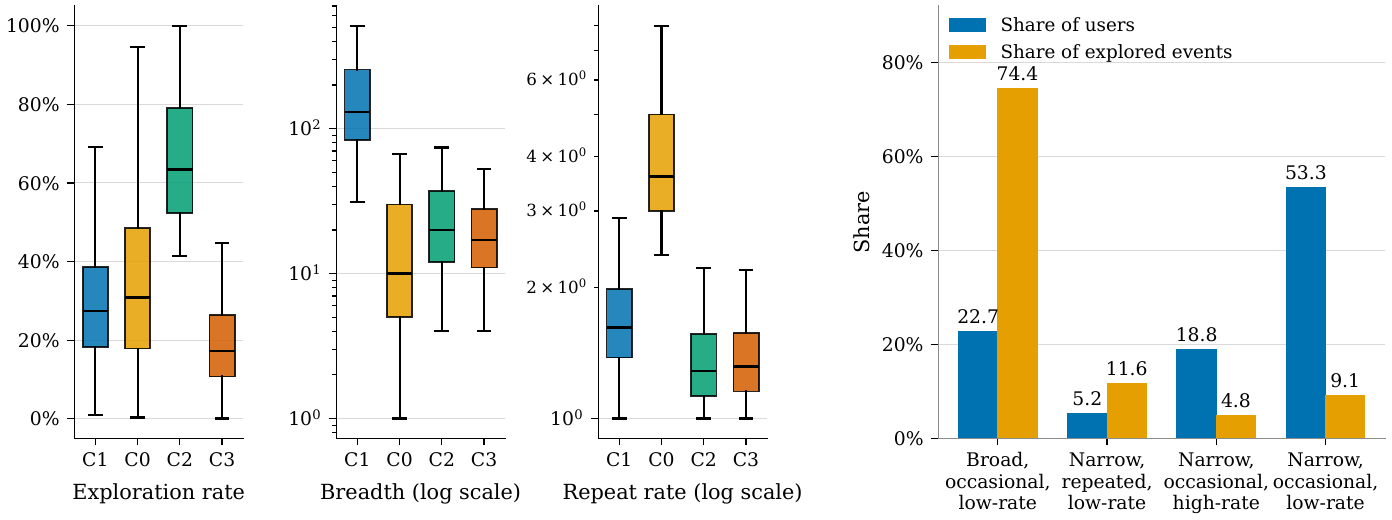}
\caption{Composition clusters ($K{=}4$): per-cluster feature distributions (exploration rate, breadth, repeat rate) and each cluster's share of users vs.\ share of explored events. Cluster ordering is the same in all panels.}
\label{fig:clusters1}
\end{figure}

\noindent\textbf{Trend-shape clustering.} A complementary view asks whether that rate itself moves with tenure, and in what direction.
Figure~\ref{fig:clusters2} shows a \emph{Downward} cluster ($34.6\%$ of users, $38.1\%\to20.7\%$ over the window), an \emph{Upward} cluster of nearly the same size ($35.0\%$, $16.3\%\to25.0\%$), and a \emph{Transient} cluster that rises then falls within the window ($30.4\%$, peaking at $26.7\%$ in week $2$ before ending at $16.9\%$).
The pooled rate over these users does decline, from $24.3\%$ at $t=0$ to $21.0\%$ at $t=7$, but that aggregate obscures real heterogeneity: a third of users are trending the other way entirely.

\begin{figure}[!ht]
\centering
\includegraphics[width=.8\linewidth]{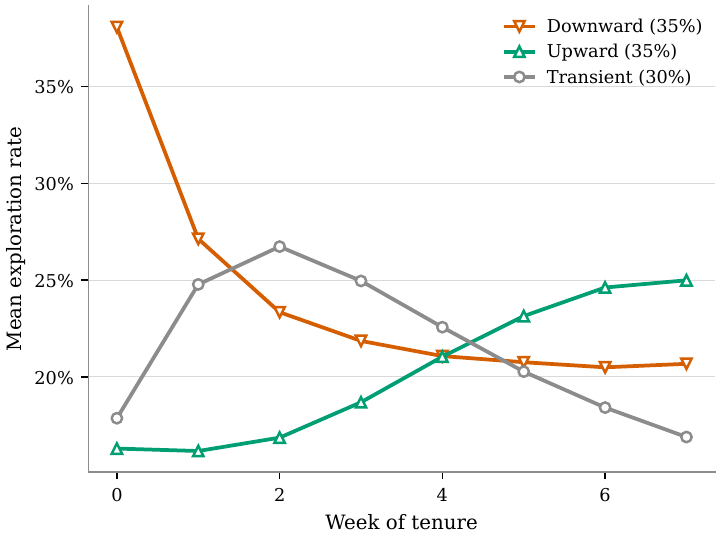}
\caption{Weekly exploration-rate trajectory clusters ($K{=}3$): mean per-cluster trajectory over the first eight weeks of tenure, share of users in parentheses.}
\label{fig:clusters2}
\end{figure}
\begin{comment}
\noindent\textbf{Crossing the two.} The two clusterings are related but not redundant.
Crossing composition against trend shape for the $103{,}290$ users both classify, the narrow, high-rate cluster is the most concentrated on a single shape ($58.8\%$ Downward, $21.9\%$ Upward, $19.2\%$ Transient), and the broad, low-repeat volume-driver cluster leans the same way more moderately ($43.7\%/29.0\%/27.3\%$).
The narrow, repeated cluster splits closer to even ($37.4\%/32.1\%/30.5\%$), and the narrow, occasional majority, the population's baseline, lands closest of all to the marginal thirds ($33.9\%/35.8\%/30.3\%$).
High-rate exploration, in other words, is disproportionately a phase rather than a stable trait: the users who explore most intensely are also the ones most likely on their way down from it.
\end{comment}

\subsection{RQ3: Does Escaped Attention Convert Into New Ties?}

Having established that a quarter of attention escapes the follow graph, we ask whether that escaped contact nonetheless produces new ties. Thirty-day conversion, pooled across the $22$M explored pairs, is $8.4\%$, with over nine tenths of conversions occurring within the first week.
This pooled rate, however, obscures a substantial dependence on the scope of first contact: pairs whose first contact remained within two hops of the user's own network convert at $11.9\%$, while pairs whose first contact lay beyond that boundary convert at $3.3\%$, a $3.6$-to-$1$ ratio (Table~\ref{tab:pairscope}). The gap persists after correcting for right-censored pairs ($12.0\%$ versus $3.4\%$ on pairs with a complete thirty-day horizon) and after adjustment for first-event type and seven user-level covariates: net of these controls, a distant first contact carries odds of conversion of $0.351$ ($95\%$ CI $0.335$--$0.368$) relative to a local one (Table~\ref{tab:model}; full model specification in Appendix~\ref{app:model}).

The remaining covariates in the adjusted model indicate that the scope effect is not simply a proxy for user activity, and in one respect runs counter to it. A standard-deviation increase in log out-degree at first contact raises the odds of conversion to $1.234$, consistent with well-connected users being more disposed to convert exploratory contact into ties. A standard-deviation increase in log total interaction volume, by contrast, lowers the odds to $0.600$: users with the highest interaction volume are, net of everything else, less likely to convert any given explored pair. Exploration rate enters negatively as well (OR $0.851$) and overall local share positively (OR $1.137$), while concentration of exploratory attention across fewer authors is indistinguishable from no effect (OR $0.998$, CI $0.980$--$1.016$). Intervals are clustered by user, which absorbs the many pairs each user contributes. Appendix~\ref{app:sensitivity} tests the scope effect against author-side confounding, heterogeneity across user strata, and redundancy with the pair's subsequent event sequence.

\begin{table}[!ht]
\centering
\small
\caption{Observed conversion by scope of first exploratory contact. ``Ever'' allows a follow at any point in the window.}
\label{tab:pairscope}
\begin{tabular}{lrrrrr}
\toprule
First scope & Explored pairs & Share & $7$d & $30$d & Ever \\
\midrule
Local two-hop & $13{,}317{,}906$ & $59.7\%$ & $10.7\%$ & $11.9\%$ & $12.4\%$ \\
Beyond two hops & $8{,}978{,}385$ & $40.3\%$ & $3.1\%$ & $3.3\%$ & $3.4\%$ \\
\midrule
All pairs & $22{,}296{,}291$ & $100.0\%$ & $7.6\%$ & $8.4\%$ & $8.8\%$ \\
\bottomrule
\end{tabular}
\end{table}

\begin{table}[!ht]
\centering
\small
\caption{Adjusted logistic model of $30$-day conversion.}
\label{tab:model}
\begin{tabular}{lrr@{\qquad}lrr}
\toprule
Term & OR & $95\%$ CI & Term & OR & $95\%$ CI \\
\midrule
First type: reply & $1.111$ & $1.042$--$1.185$ & $z$ log out-degree & $1.234$ & $1.144$--$1.331$ \\
First type: repost & $0.810$ & $0.680$--$0.965$ & $z$ log in-degree & $1.157$ & $1.077$--$1.243$ \\
First type: quote & $0.691$ & $0.603$--$0.791$ & $z$ observed tenure & $0.718$ & $0.678$--$0.760$ \\
First scope: distant & $0.351$ & $0.335$--$0.368$ & $z$ exploration rate & $0.851$ & $0.817$--$0.887$ \\
$z$ log total interactions & $0.600$ & $0.551$--$0.654$ & $z$ concentration & $0.998$ & $0.980$--$1.016$ \\
 & & & $z$ local share & $1.137$ & $1.096$--$1.180$ \\
\bottomrule
\end{tabular}
\end{table}

\section{Discussion and Conclusion}\label{sec:discussion}
Our results show that attention on Bluesky is strongly anchored to the social network users have already constructed. Although one quarter of interactions escape the follow graph, most remain structurally nearby: $80.6\%$ of all interactions land on an author within two follow hops, and a further $12.2\%$ reach the user by reshare from an account they follow. The large gap from the expected-degree null --- $80.8\%$ against $22.4\%$ on the scored interactions --- further shows that this concentration is not a trivial consequence of degree structure. In a platform where the follow graph constitutes a meaningful baseline for content access, network proximity remains a powerful correlate of where interaction occurs.
\\ \ \\
Exploration beyond this neighborhood is nevertheless substantial and highly uneven. A relatively small group of broad-reaching users generates most exploratory activity, while longitudinal patterns show that exploration is not a single stable behavioral tendency: declining, increasing, and transient trajectories coexist. The aggregate decline with tenure therefore conceals considerable individual heterogeneity. More importantly, structural distance appears consequential for network evolution. Contacts beyond two hops convert into follows at only $3.3\%$, compared with $11.9\%$ for local exploratory contacts, and the difference remains after adjustment for observed user characteristics. Distant attention is therefore not simply less common; it is also less likely to become incorporated into the user's subsequent social network.
\\ \ \\
Concluding, these findings suggest a distinction between \emph{exploration as exposure} and \emph{exploration as structural change}.
Users frequently interact beyond accounts they already follow, but most such activity remains close to the existing graph, and encounters that travel further rarely produce new ties. This qualifies the intuitive idea that boundary-spanning attention necessarily expands users' social neighborhoods: on Bluesky, exploration is substantial in volume, but its capacity to reshape
the follow graph appears limited.
\\ \ \\
These conclusions remain subject to several limitations. Interaction records capture acted-upon attention rather than exposure itself, two-hop locality is reconstructed from the available follow graph rather than from exact delivery paths, and the conversion models are observational and do not establish a causal effect of network distance. The study also covers an early period of Bluesky's growth, so the magnitude of these patterns may change as the platform and its recommendation ecosystem evolve. Future work should test whether the same relationship between network proximity, exploration, and tie formation holds in later periods and on platforms where algorithmic ranking plays a more central role.

\appendix
\section{Robustness, Validation, and Model Specifications}

\noindent{\textbf{Expected-degree null.}}\label{app:null} \sloppy We use a an expected-degree Chung--Lu model: each ordered pair $(u,a)$ is taken to be connected independently with probability $k_{\mathrm{out}}(u)k_{\mathrm{in}}(a)/M$, and the expected two-hop path count is scaled by $S_2=\sum_m k_{\mathrm{in}}(m)k_{\mathrm{out}}(m)$. The null is evaluated honoring each scored interaction's own timestamp. 
Because the null is a statement about follow-path distance, the observed quantity compared against it is the within-two-hop share (the followed and two-hop-local tiers) and excludes the relay tier, which is an information route rather than a follow path. 
{A separate check on three time-separated cohorts, drawn early, mid, and late in the labelled log and scored against the same null, indicates the gap is not a period artifact: the within-two-hop share is $82.0\%$, $78.8\%$ and $80.3\%$ against nulls of $28.2\%$, $32.5\%$ and $28.0\%$, on $6.20$M, $1.63$M and $1.69$M interactions respectively.}

\noindent\textbf{Clustering method.}\label{app:clustering} Both clusterings use $K$-means with $K$ chosen by silhouette (scanned $K=2$--$10$, evaluated on a $100$k-user subsample per $K$), a GMM at the selected $K$ as a cross-model check, and Hungarian-matched split-half ARI for stability.
Composition features: exploration rate, $\log(1{+}\text{breadth})$, $\log(1{+}\text{events}/\text{breadth})$, floored at $\geq10$ explored events, below which breadth and repeat rate are near-deterministic.
Trajectory features: each user's eight weekly exploration rates (tenure week $=(\text{ts}-\text{ts}_{\min})\,\text{div}\,604{,}800$ from their own first observed interaction), smoothed toward the window mean with an additive constant of five, for users observed the complete window with $\geq20$ interactions in it; the window length trades population size against tenure covered ($384$k users qualify at four weeks, $55$k at fourteen).
For the trend-shape clustering, each user's eight-value vector is additionally $z$-scored (own mean and standard deviation removed) before clustering, so the fitted shape reflects the pattern of change rather than its level; $K$ is scanned $2$--$10$ as for the composition clustering.
\noindent\textit{Composition clustering ($K=4$).} Silhouette peaks cleanly at $K=4$ ($0.376$), clearly ahead of every other $K$ scanned from $2$ to $10$, and a two-component PCA of the three features captures roughly two-thirds of their variance. The partition is highly reproducible under resampling (split-half ARI $0.984$), though only moderately so across algorithms (KMeans versus a GMM at the same $K$, ARI $0.441$), which we read as a stable discretization of the space.
\noindent\textit{Trend-shape clustering ($K=3$).} Here the silhouette curve is comparatively flat: $K=2$ is nominally best, but every larger $K$ up to $10$ lands in a similarly narrow, unremarkable band, with no real second peak. We report $K=3$ over the silhouette-optimal $K=2$ because it agrees far better across algorithms (ARI $0.509$ vs.\ $0.226$) while remaining just as reproducible under resampling (split-half ARI $0.956$, against $0.968$ for $K=2$).

\noindent{\textbf{Conversion model specification.}}\label{app:model} {The regression of Section~\ref{sec:results} (RQ3) is a pair-level logistic model of $30$-day conversion --- a binary indicator for a follow-creation record by the target author within thirty days of the pair's first exploratory contact $t_0$.
Every user covariate is computed from events strictly preceding $t_0$ to avoid leakage. The model therefore conditions on the user's state at the moment of first contact.
The specification includes indicator terms for the pair's first-event type (reply, repost, quote; like as reference) and for distant first scope (two-hop local as reference), together with the seven user covariates of Table~\ref{tab:covariates}. All continuous covariates enter as $\log(1{+}x)$ transforms, standardized within the fitted sample; reported odds ratios are exponentiated coefficients at that standardization.
Confidence intervals are clustered by user, which accounts for the many pairs a single user contributes. They do not account for authors shared across users.}

\begin{table}[!ht]
\centering
\small
\caption{{User covariates used in the RQ3 conversion model (Table~\ref{tab:model}), each computed from the user's events strictly preceding the pair's first exploratory contact $t_0$.}}
\label{tab:covariates}
\begin{tabular}{@{}lp{0.6\linewidth}@{}}
\toprule
Covariate & Definition \\
\midrule
Total interaction volume & {Total interactions by the user before $t_0$.} \\
Out-degree & Number of distinct accounts followed as of $t_0$. \\
In-degree & {Number of distinct accounts following the user as of $t_0$.} \\
Observed tenure & {Days between the user's first observed interaction and $t_0$.} \\
Exploration rate & {Share of the user's interactions before $t_0$ classified as exploratory (out-of-network, non-relay-mediated).} \\
Concentration & Herfindahl--Hirschman index of the user's exploratory interactions across explored authors before $t_0$ ($\sum_i p_i^2$, with $p_i$ the share of exploratory interactions directed at explored author $i$); higher values indicate attention concentrated on fewer authors. \\
Local share & {Share of the user's exploratory interactions before $t_0$ that are two-hop local rather than distant.} \\
\bottomrule
\end{tabular}
\end{table}

\noindent\textbf{Sensitivity.}\label{app:sensitivity}
There are three ways the scope effect could be an artifact.
The first is confounding by the target: distant authors might simply be harder to follow. Adding the author's in-degree, out-degree and received-interaction volume as of $t_0$, each estimated on the sampled side of the collection, moves the odds ratio from $0.351$ to $0.289$. Author-side connectivity therefore does not explain away the association between structural proximity and subsequent tie formation.
The second is that the effect belongs to one kind of user: someone who follows few accounts has a small two-hop neighbourhood, so almost everything they touch counts as distant by construction. Refitting on fifteen overlapping subsets, by activity, degree, tenure, exploration rate and position in the observation window, leaves the odds ratio between $0.331$ and $0.380$, every interval excluding one. The penalty on distant contact is therefore a regularity of tie formation across the population, not an artifact of users whose neighbourhoods are too small to reach locally.
The third is that scope only stands in for how the pair developed: distant pairs might convert less simply because they never interacted again. Replacing the indicator with five measures of the pair's own event sequence over the thirty days after $t_0$ --- its volume, its spread across days, its composition by event type, and the share of those events that were exploratory and two-hop local --- puts that local share at $1.682$ across its full range, well short of the $1/0.351=2.849$ redundancy would imply. The two track each other without coinciding: among pairs contacted again, the share runs $0.834$ for local-first pairs against $0.073$ for distant-first ones. 

\bibliographystyle{spmpsci}
\bibliography{refs}

\end{document}